# Is non-Gaussianity sufficient to produce long-range volatile correlations?

Radhakrishnan Nagarajan

## Abstract

Scaling analysis of the magnitude series (volatile series) has been proposed recently to identify possible nonlinear/multifractal signatures in the given data [1-3]. In this letter, correlations of volatile series generated from stationary first-order linear feedback process with Gaussian and non-Gaussian innovations are investigated. While volatile correlations corresponding to Gaussian innovations exhibited uncorrelated behavior across all time scales, those of non-Gaussian innovations showed significant deviation from uncorrelated behavior even at large time scales. The results presented raise the intriguing question whether non-Gaussian innovations can be sufficient to realize long-range volatile correlations.

**Author for Correspondence**

Radhakrishnan Nagarajan
Center on Aging, UAMS
629 Jack Stephens Drive, Room: 3105
Little Rock, AR 72205, USA
Email: nagarajanradhakrish@uams.edu

## 1. Introduction

Detrended fluctuation analysis (DFA) [4] and its extensions [1-3, 5-6] have been used widely to determine the nature of correlations in synthetic and experimental data obtained from a wide-range of complex systems. Recently [1-3], analysis of the magnitude series of the given empirical sample have been used to gain further insight into the underlying dynamics [5-10]. More importantly, long-range correlation in the magnitude series was found to be indicative of nonlinear and possibly multifractal signatures in the given data [1-3, 5-10]. Several models have also been proposed recently to generate volatile correlations under certain constraints [11]. In the present study, we investigate the impact of Gaussian and non-Gaussian innovations on the scaling of magnitude series generated from stationary first order linear feedback processes.

## 2. Methods

First-order linear feedback process represents the most elementary of the stochastic processes, and is given by the expression

$$x_n = \phi\, x_{n-1} + \in_n \qquad (1)$$

where $\in_n$ represents identical and independently distributed (i.i.d) process (white noise) sampled from a given distribution, also known as *innovations*. The $x_n^{th}$ sample is related to the $x_{n-1}^{th}$ sample through the process parameter $\phi$. For the same reason, processes such as (1) are termed as finite-memory or Markov processes. Each $x_n$ is a weighted sum or linear combination of innovations $\in_n$. Therefore, (1) is a *linearly correlated noise* whose distribution is governed by the $\in_n$. It can be shown analytically that the above process (1) is stationary for $|\phi| < 1$ with associated auto-correlation function $r(k) = \phi^k$ (see Appendix). In the present study, we considered process parameters $\phi$ = 0.95 and $\phi$ = 0.65. Unlike the latter, $\phi$ = 0.95 is close to the non-stationary regime $\phi$ = 1, resulting in slow decay of the auto-correlation function. We investigate the scaling behavior of (1) with Gaussian as well as non-Gaussian innovations $\in_n$ sampled from five different distributions, namely

**NORM**: zero-mean unit variance innovations $\in_n$, sampled from a Gaussian distributed white noise ($g$) with probability density function $f(g) = \frac{1}{\sqrt{2p}} e^{-x^2/2}, x \in (-\infty, \infty)$.

**SQNORM**: zero-mean unit variance innovations $\in_n$, sampled from squared transform of a Gaussian distributed white noise, i.e. $g_1 = g^2$.

**EXPNORM**: zero-mean unit variance innovations $\in_n$, sampled from exponential transform of a Gaussian distributed white noise i.e. $g_2 = e^g$.

**UNI**: zero-mean unit variance innovations $\in_n$, sampled from uniformly distributed white noise ($u$) with probability density function $f(u) = 1/(b-a), u \in (a, b)$.

**LOGUNI**: zero-mean unit variance innovations $\in_n$, sampled from negative log transform of a uniformly distributed white noise i.e. $u_1 = -log(u)$.

The above abbreviations shall be used in the subsequent sections. It can be shown analytically that second order moments are sufficient to completely describe first order linear feedback process with Gaussian innovations. However, this is not true in the case of non-Gaussian innovations where higher order statistics are required to sufficiently describe the process. Two popular statistics used in literature to reflect the deviation from Gaussianity are skewness $(y)$ and kurtosis $(k)$. Skewness $(y)$ and kurtosis $(k)$ of the innovations (NORM, EXPNORM, SQNORM, UNI and LOGUNI) are shown in Figs. 1(a-e) respectively. Those of their corresponding linear feedback processes with parameters ($q$ = 0.95, N = $2^{16}$) are shown in Figs. 1(f-j) respectively. While NORM and UNI are symmetric distributions $(y = 0)$, SQNORM,

EXPNORM and LOGUNI are asymmetric $(\gamma \neq 0)$. Kurtosis of UNI $(k = 1.8)$ and NORM $(k = 3)$ are dissimilar, Fig. (1a and 1d), however, those of their corresponding linear feedback processes are similar $(k = 3)$, Fig. (1f and 1i). As shall be shown later volatile correlations of (1), with UNI showed minimal discrepancy from those with NORM.

## 3. Results

Classical power-spectral analysis is used widely to investigate correlations is stationary linear processes such as (1). Power-spectrum of a stationary process is related to its auto-correlation function by the Wiener-Khinchin theorem. As noted earlier (Appendix), the expression of the auto-correlation for the first-order linear feedback process $x_n$ (1) is governed solely by the process parameter $q$ and is immune to the distribution of the innovations $\in_n$. Thus it might not be surprising to note that first-order linear feedback processes with Gaussian (NORM) and non-Gaussian (SQNORM, EXPNORM, UNI and LOGUNI) innovations revealed similar spectral signatures, Fig. 2a, also reflected in the scaling of their fluctuation function F(s) with time scale (s), Fig. 3a, obtained using DFA with fourth order polynomial detrending. In literature, volatility series of the given data has been generated using continuous and discontinuous, static, memoryless nonlinear transforms [3]. The power spectrum of the magnitude series generated from mean subtracted $x$ (1) with transforms $|x|$, $x^2$, $|x_n - x_{n-1}|$, and innovations NORM, SQNORM, EXPNORM, UNI and LOGUNI is shown in Fig. 2. As expected the qualitative behavior of the power-spectrum, hence the correlation showed a marked change across the three different transforms [12]. More importantly, the transforms $|x|$, $x^2$ do not exhibit considerable variation across the various innovations, Figs, 2b and 2c. Also reflected in the scaling of their fluctuation function F(s) with respect to the time scale (s), Figs. 3b and 3c, obtained using DFA with fourth order polynomial detrending [2, 5]. However, the power-spectrum, Fig. 2d, and the scaling of the fluctuation function, Fig. 3d, of the volatility series generated using the transform $|x_n - x_{n-1}|$ exhibits a significant change across the various types of innovations. More importantly, power-spectrum of the volatility series with non-Gaussian innovations (SQNORM, EXPNORM, LOGUNI) exhibited dominant low-frequency characteristic of long-range correlated noise, unlike their Gaussian (NORM) counterpart, Fig. 2d. As noted earlier, volatility scaling with innovations UNI was uncorrelated ($\alpha = 0.5$) similar to that of NORM across all time scales. This has to be contrasted with those of (SQNORM, EXPNORM and LOGUNI) which exhibited significant deviation from ($\alpha = 0.5$) across a wide-range of time scales, Figs. 2d and 3d. As noted from Figs. 1c and 2c, the discontinuous, static, memoryless nonlinear transform $|y|$ singularly fails to impose uncorrelated structure on (1) with Gaussian innovations across all time scales, Figs. 1b and 2b. It can be shown analytically that differencing the process (1) is immune to the innovations. Thus the observed discrepancy in the scaling across innovations (NORM, SQNORM, EXPNORM and LOGUNI) can be attributed to the specific-form of the volatility transform, i.e. a high pass filter $(x_n - x_{n-1})$ in conjunction with the discontinuous, static, memoryless nonlinear transform $|y|$. In the subsequent discussions volatility transform implicitly refers to the transform $|x_n - x_{n-1}|$.

Prior to investigating the scaling of the volatile series, we investigated the scaling of the linear feedback process (1) with Gaussian (NORM) and non-Gaussian innovations (SQNORM, EXPNORM, UNI, and LOGUNI) with lengths ($N = 2^{12}$, $2^{16}$, $2^{20}$) and parameters ($q = 0.95$ and 0.65). The log-log plot of F(s) versus (s) exhibited significant nonlinearity with characteristic crossovers, Fig. 3a. In such cases, scaling exponent estimation by linear regression of F(s) versus (s) over the entire range of time scales can lead to spurious conclusions. In order to circumvent these issues we estimated the local scaling exponent, by linear regression of overlapping moving windows. This was accomplished by choosing a window containing five points, estimate the exponent by local linear regression of the points in that window, shift the window by two points and repeat the procedure. Thus as a result, we obtain the scaling exponents α(s) as a function of the time scales (s). The local scaling exponents of the (1) with the five different innovations is shown in Fig. 4. The most important observation from the context of the present study is that the scaling of (1) is immune to the choice of the innovations. As shall be shown later this is not necessarily true for the corresponding volatility series. The similarity in the scaling can be attributed to the fact that the correlation function of (1) is independent of the innovations (Appendix), also reflected in the similarity of the power-spectrum, Fig. 2a. For ($q$ = 0.95, N = $2^{12}$), Fig. 4a, one fails to see the transition from correlated to uncorrelated regime. Thus such a small sample size can lead to spurious conclusion regarding the dynamics. For ($q$ = 0.95, N = $2^{16}$) the local scaling exponent does exhibit a transition from correlated to uncorrelated regime at time scale (s ~ $2^{12}$), Fig. 4b. However, the number of points in the uncorrelated regime is less preventing reliable estimation of the scaling exponents. For ($q$ = 0.95, N = $2^{16}$) there is a clear transition from correlated to uncorrelated regime (s ~ $2^{13}$), Fig 4c. Unlike the case of ($q$ = 0.95), transition from correlated to uncorrelated regime can be observed even at small sample sizes such as, N = $2^{12}$, for ($q$ = 0.65), Figs. 4d, 4e and 4f. Alternatively, the uncorrelated regime dominates the scaling region for smaller values of $q$, this is to be expected as correlation decays faster for smaller $q$ (Appendix). Thus finite size effects play an important role in the interpretation of the scaling behavior, especially as the process parameter tends to unity, $q \to 1$.

A similar analysis was carried out for the volatility series $|x_n - x_{n-1}|$ of the linear feedback process (1) with innovations ($\in_n$ : NORM, SQNORM, EXPNORM, UNI and LOGUNI), lengths (N = $2^{12}$, $2^{16}$, $2^{20}$) and process parameter ($q$ = 0.95 and 0.65). The scaling of the volatility series with Gaussian innovations (NORM), which exhibited uncorrelated behavior across all time scales, was used as an internal control. Twenty five independent realizations were generated for each case in order to estimate the uncertainty in the scaling exponent estimate across the time scales. The average and the standard deviation of the local volatility scaling exponents, $\alpha^v(s)$, for the non-Gaussian innovations are enclosed in Figs. 5 and 6. Those obtained with Gaussian innovations with the respective parameters $q$ and lengths N is enclosed in each of the subplots as reference. The scaling exponent corresponding to uncorrelated noise ($\alpha = 0.5$) is also shown by the horizontal solid lines as a reference. For parameter ($q = 0.95$), the average $\alpha^v(s)$ corresponding to Gaussian (NORM) and uniformly distributed (UNI) innovations resemble that of uncorrelated noise irrespective of the choice of the length ($N = 2^{12}, 2^{16}, 2^{20}$), Figs. 5c, 5g and 5k. However, the average $\alpha^v(s)$ for (SQNORM, EXPNORM and LOGUNI) exhibited considerable deviation from those of uncorrelated noise across the entire length of the time scales ($s \sim 2^6$ to $2^9$) for ($N = 2^{12}$), Figs. 5a, 5b and 5d. The uncertainty in the $\alpha^v(s)$ estimate as reflected by the standard deviation was also maximum at this length ($N = 2^{12}$). For ($N = 2^{16}$), average $\alpha^v(s)$ estimate for non-Gaussian innovations (SQNORM, EXPNORM and LOGUNI) exhibited significant discrepancies from their Gaussian counterpart across a majority of the time scales ($s \sim 2^6$ to $2^{11}$), Figs. 5e, 5f and 5h. For time scales ($s > 2^{11}$) average $\alpha^v(s)$ estimate converged to those of uncorrelated behavior. However, the uncertainty is high preventing reliable estimation of the scaling exponents at time scales corresponding to the uncorrelated regimes. Such a behavior was consistent across all the innovations, Figs. 5e-5h. As in the case of ($N = 2^{16}$), average $\alpha^v(s)$ estimate for non-Gaussian innovations for ($N = 2^{20}$) showed a significant deviation from uncorrelated behavior across a wide-range of time scales ($s \sim 2^6$ to $2^{12}$) Figs. 5i, 5j and 5l. However, the average $\alpha^v(s)$ estimate converged to that of uncorrelated behavior for ($s > 2^{12}$). A similar analysis was carried out for the volatility series of (1) with parameter ($q = 0.65$), with innovations (NORM, SQNORM, EXPNORM, UNI and LOGUNI) and lengths ($N = 2^{12}, 2^{16}, 2^{20}$), Fig. 6. As with the case of ($q = 0.95$), $\alpha^v(s)$ estimates of NORM and UNI were similar to that of uncorrelated noise, Figs. 6c, 6g and 6k. As observed earlier, $\alpha^v(s)$ estimates exhibited maximum uncertainty for ($N = 2^{12}$) across the five different innovations, which diminished with increasing length ($N = 2^{16}, 2^{20}$). The $\alpha^{v}(s)$ estimates for non-Gaussian innovations (SQNORM, EXPNORM and LOGUNI) exhibited a marked crossover from correlated behavior to uncorrelated behavior unlike their Gaussian counterpart, Fig. 6. Unlike the case of ($q = 0.95$), the average $\alpha^{v}(s)$ estimates across the various innovations for ($q = 0.65$) was dominated by uncorrelated behavior similar to that of their Gaussian (NORM) counterpart.

## 4. Discussion

Previous studies have demonstrated that volatility transform $|x_n - x_{n-1}|$ renders the scaling of monofractal noise to be uncorrelated ($\alpha^{v} = 0.5$). A significant deviation from ($\alpha^{v} = 0.5$) have been attributed to possible nonlinear/multifractal signatures in the given data. In the present study, we investigated the impact of Gaussian (NORM) and non-Gaussian (SQNORM, EXPNORM, UNI and LOGUNI) innovations on the volatility scaling of stationary markovian first order linear feedback process. While the scaling of the linear feedback process was immune to the choice of the innovations, those of their volatility series were not. Volatile correlations of the linear feedback process with uniformly distributed innovations were uncorrelated across all time scales similar to those of their Gaussian (NORM) counterpart. However, those of (SQNORM, EXPNORM and LOGUNI) showed significant correlation across a majority of the time scales unlike those of NORM. The impact of non-Gaussianity was prominent for process parameter $q$ close to unity. The present study also investigated the effect of sample-sizes on the local scaling exponent estimates. The scaling region was dominated by correlated behavior for sample sizes ($N < 2^{12}$). The transition from correlated to uncorrelated behavior in the case of non-Guassian innovations was unclear with maximum uncertainty at larger time scales for sample sizes ($N < 2^{16}$). The results presented clearly demonstrate that non-Gausssian innovations introduce volatile correlations even at large time scales unlike their Gaussian counterpart.

Therefore, the present study raises the intriguing question whether non-Gaussianity may be sufficient to produce volatile correlations. In experimental data sets, one often has access to a single realization of the process. Under ergodic assumptions, this single realization can be thought to be representative of the underlying dynamics. However, as shown in the present study a sufficiently long data might be necessary in order avoid spurious conclusion of long-range volatile correlation.

**Acknowledgements**

I would like to thank the reviewers for helpful comments and suggestions which were useful in enhancing the quality of the manuscript. The present study is supported by funds from National Library of Medicine (1R03LM008853-1) and junior faculty grant from American Federation for Aging Research (AFAR).

**Appendix**

Consider the first-order linear feedback process $x_n = q\, x_{n-1} + \in_n$, where $\in_n$ are i.i.d innovations sampled from a white noise with a given distribution. The correlation between samples $x_n$ and $x_{n-k}$ is given by the expression

$$R(k) = E(x_n x_{n-k})\text{, where } E \text{ represents the expectation operator}$$

For $k$ = 1, we have

$$R(1) = E(x_n x_{n-1}) = E(\ (q\, x_{n-1} + \in_n) x_{n-1})$$
$$= q\, E(x_{n-1} x_{n-1}) = q\, R(0) \text{ as } E(x_{n-i} \in_n) = 0 \text{ for } i > 0$$

Similarly, for $k$ = 2, we have

$$R(2) = E(x_n x_{n-2}) = E(\ (q\, x_{n-1} + \in_n) x_{n-2}) = q\, E(x_{n-1} x_{n-2}) = q\, R(1) = q^2 R(0)$$

For the $m^{th}$ lag, we have

$$R(m) = q^m R(0)$$

The corresponding auto-correlation function is given by

$$r(m) = \frac{R(m)}{R(0)} = q^m \text{ which is stationary for } |q| < 1.$$

It is important to note that the expression for the auto-correlation function $r(m) = q^m$ and the constraints on stationarity $|q| < 1$ are immune to the distribution of $\in_n$.

**Reference**

1. Y. Ashkenazy et al. Phys. Rev. Lett. 86 (2001), 1900.
2. Y. Ashkenazy et al. Physica A. 323 (2003), 19.
3. T. Kalisky Y. Ashkenazy, and S. Havlin. Phys. Rev. E 72 (2005), 011913.
4. C.-K. Peng et al. Nature 356 (1992), 168.
5. J.W. Kantelhardt et al. Physica A 295 (2001), 441.
6. J.W. Kantelhardt et al. Physica A 316 (2002), 87.
7. J. W. Kantelhardt et al. Phys. Rev. E. 65 (2002) 051908.
8. J. W. Kantelhardt et al. Europhys. Lett. 62 (2003) 147.
9. P. Ch. Ivanov et al. Phys. Rev. E. 69 (2004) 056107.
10. K. Hu et al. Physica A. 337 (2004) 307-318.
11. B. Podobnik et al. Phys. Rev. E. 72 (2005) 026121.
12. Z. Chen et al. Phys. Rev. E. 71(2005) 011104.

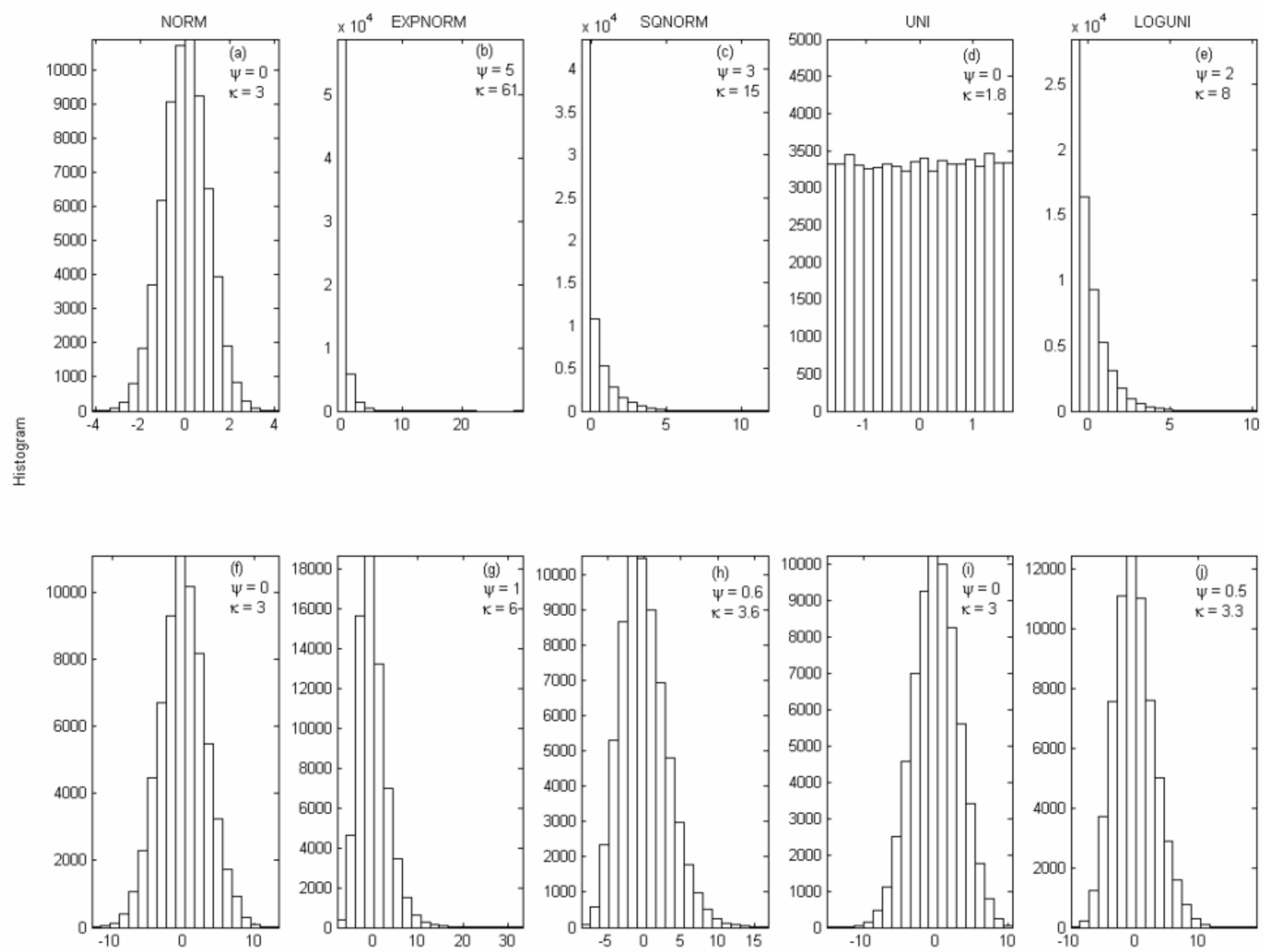

**Figure 1** Histogram of Gaussian and non-Gaussian innovations (N = $2^{16}$) namely: NORM, EXPNORM, SQNORM, UNI and LOGUNI along with their skewness $(y)$ and kurtosis $(k)$ are shown in a-e respectively. Histogram along with $(y, k)$ for the stationary first order linear feedback processes ($q$ = 0.95) generated with the above innovations is shown right below them in f-j respectively.

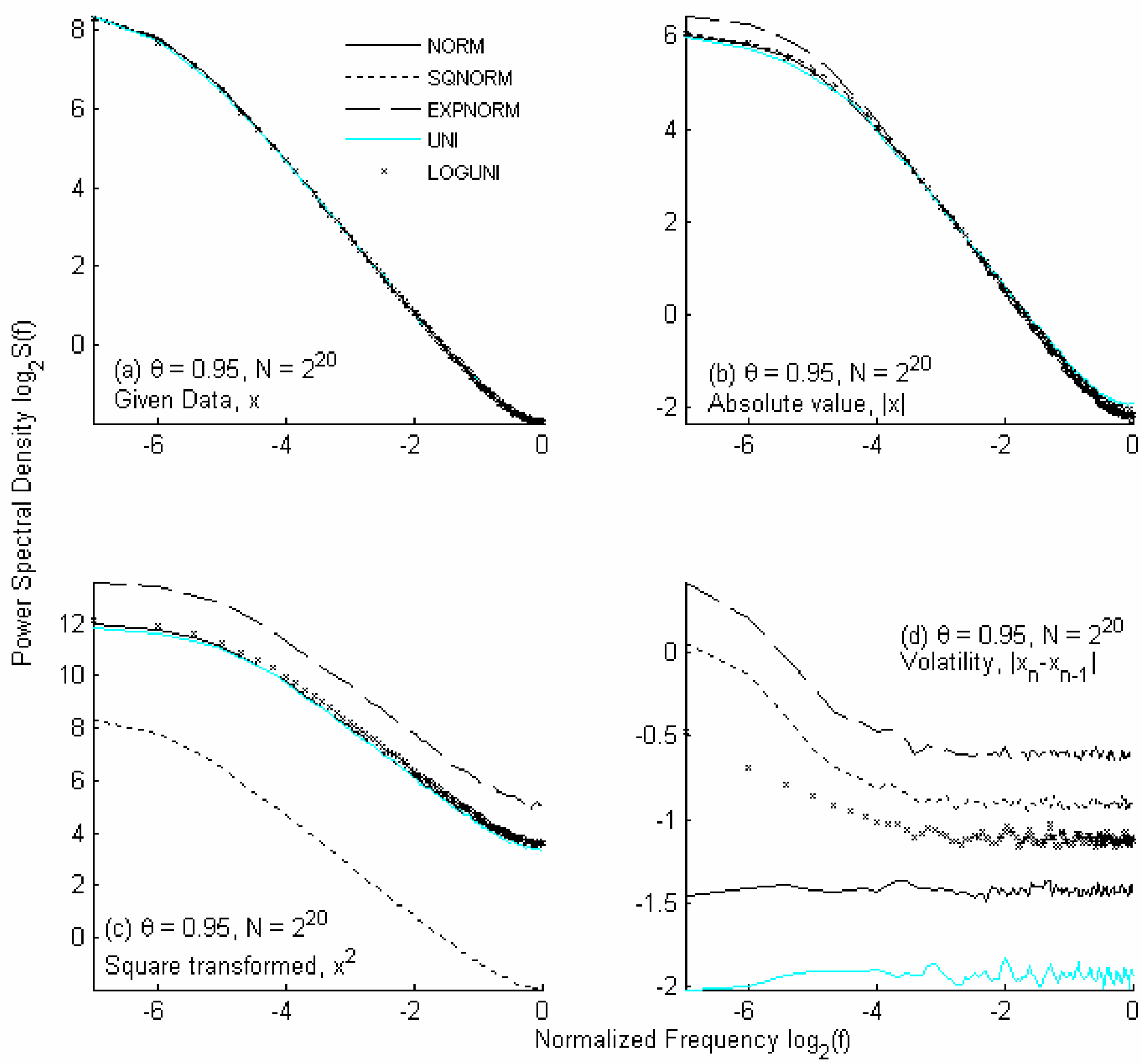

**Figure 2** Power-spectral density, $\log_2 S(f)$ versus $\log_2(f)$, for the stationary first linear feedback process ($q = 0.95$, $N = 2^{20}$) with Gaussian (NORM) and non-Gaussian innovations (SQNORM, EXPNORM, UNI, LOGUNI) is shown in (a). Power-spectral density of their corresponding nonlinear transformed counterparts $|x|$, $x^2$, $|x_n - x_{n-1}|$ is shown in (b), (c) and (d) respectively. The legend for the five different cases (NORM, SQNORM, EXPNORM, UNI and LOGUNI) is enclosed in (a).

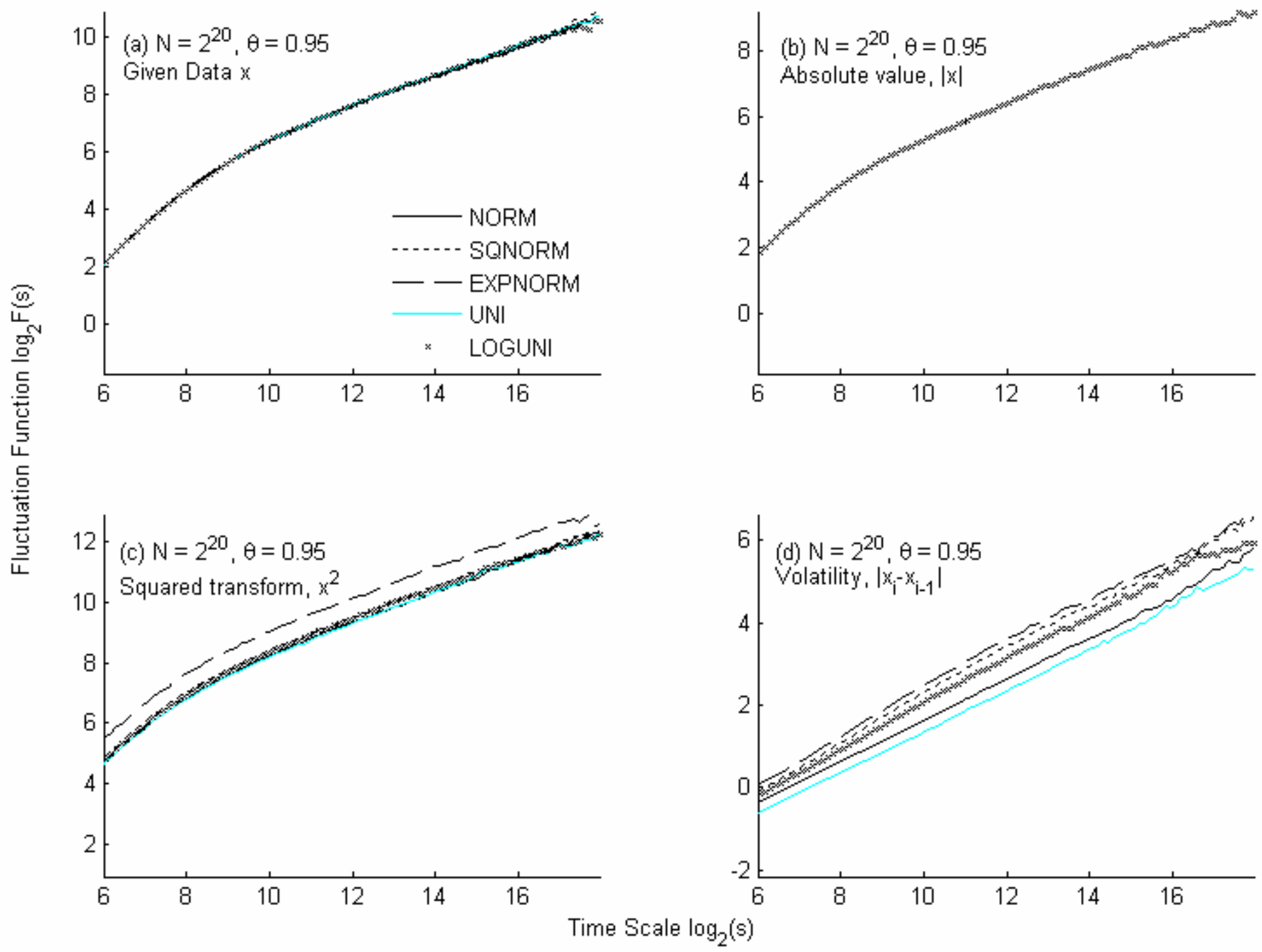

**Figure 3** Plot of the fluctuation function $\log_2 F(s)$ versus time scale $\log_2(s)$ for the first linear feedback process $(q = 0.95, \mathrm{N}=2^{20})$ with Gaussian (NORM) and non-Gaussian innovations (SQNORM, EXPNORM, UNI, LOGUNI) obtained using DFA with fourth order polynomial detrending is shown in (a). Those of their nonlinear transformed counterparts $|x|$, $x^2$, $|x_n - x_{n-1}|$ are shown in (b), (c) and (d) respectively. The legend for the five different cases (NORM, SQNORM, EXPNORM, UNI and LOGUNI) is enclosed in (a).

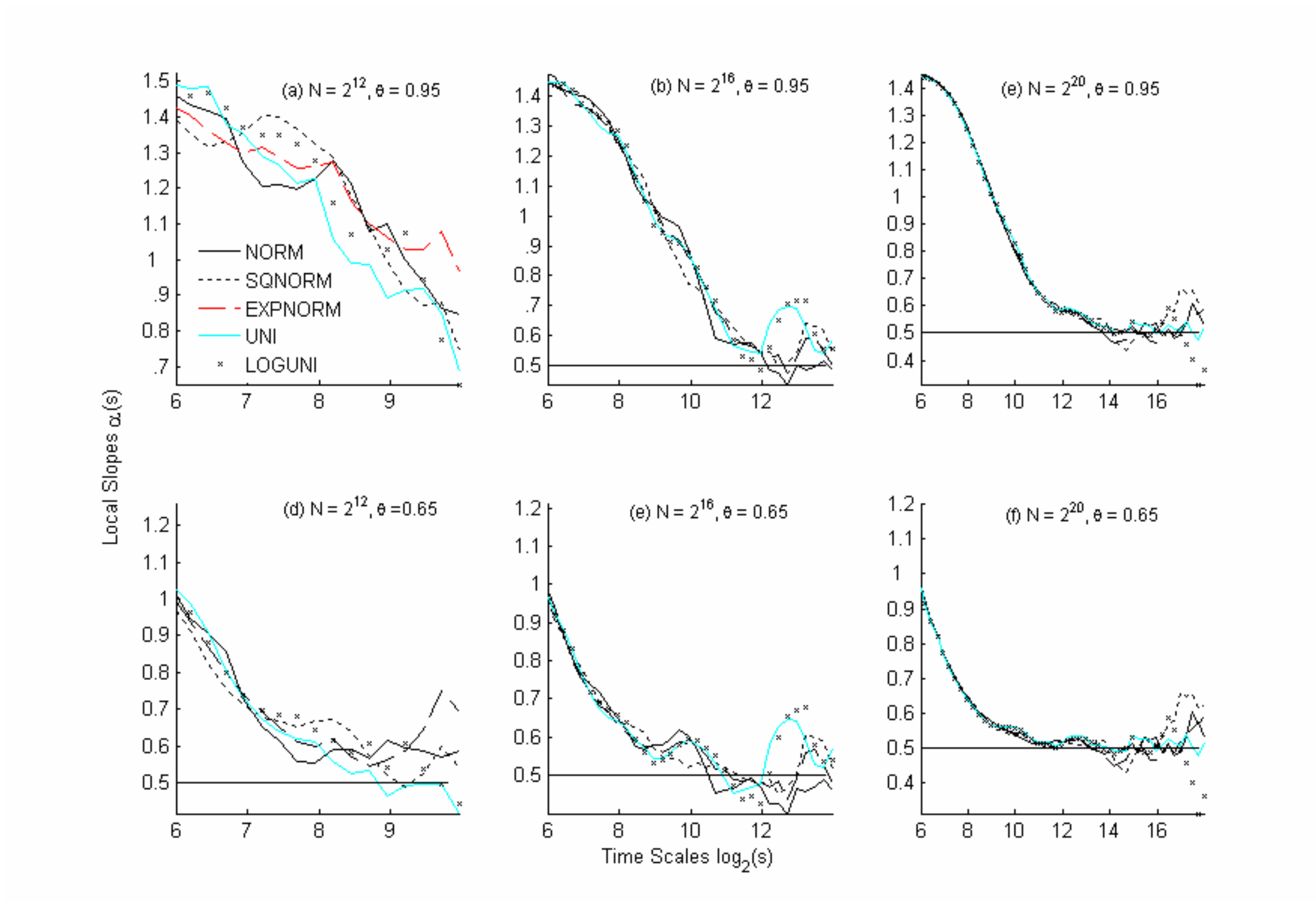

**Figure 4** Local slopes α(s) versus time scale $\log_2(s)$ for the linear feedback process (*q* = 0.95, *q* = 0.65) with Gaussian (NORM) and non-Gaussian innovations (SQNORM, EXPNORM, UNI, LOGUNI) with varying sample sizes (N = $2^{12}$, $2^{16}$ and $2^{20}$) obtained using DFA with fourth order polynomial detrending. The legend for the five different cases (NORM, SQNORM, EXPNORM, UNI and LOGUNI) is enclosed in (a). The horizontal reference line corresponds to scaling exponent (α = 0.5) characteristic of uncorrelated noise.

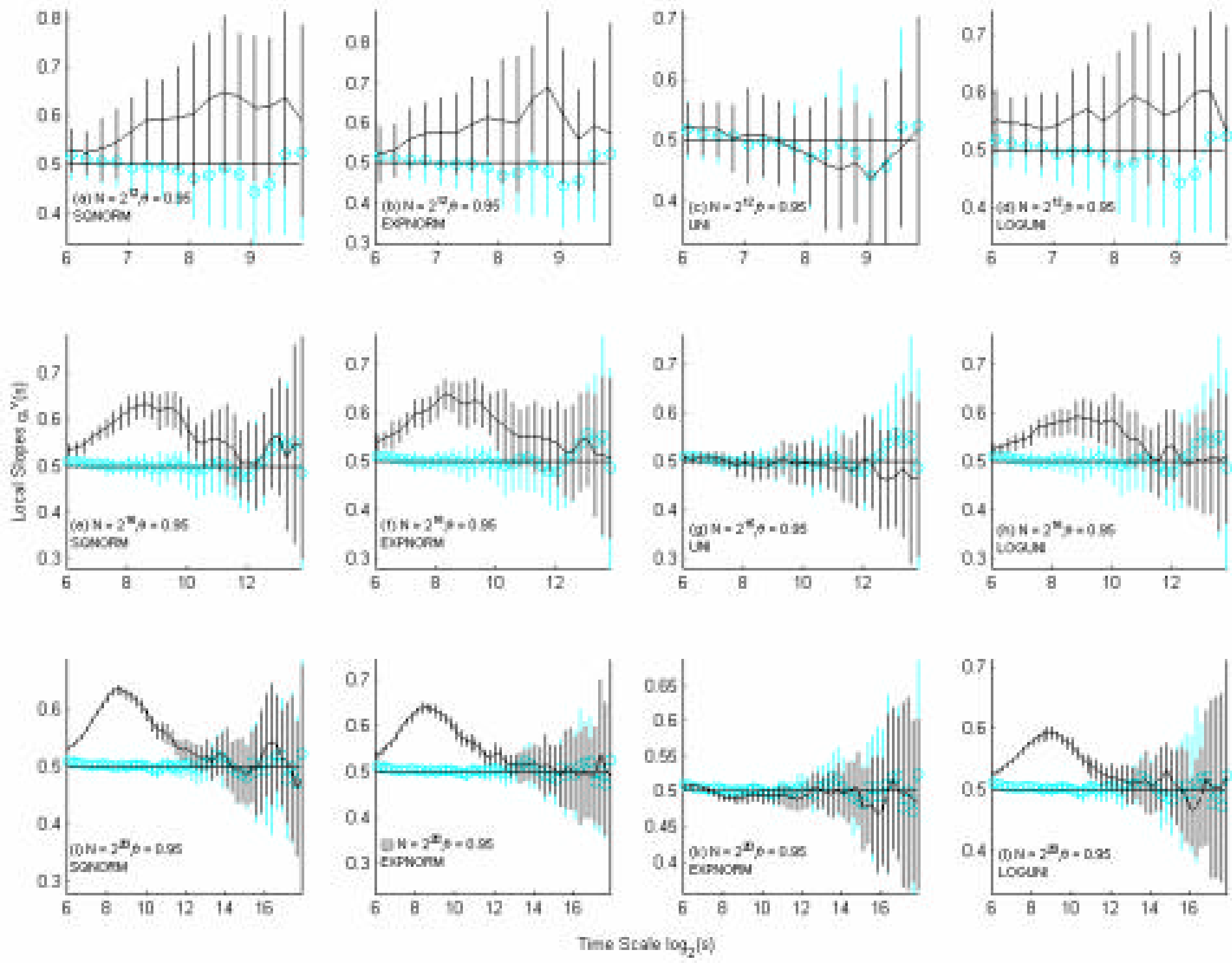

**Figure 5** Local slopes α(s) versus time scale $\log_2(s)$ of the volatility series of the stationary linear feedback process ($q$ = 0.95) with non-Gaussian innovations (SQNORM, EXPNORM, UNI, LOGUNI) and varying sample sizes (N = $2^{12}$, $2^{16}$ and $2^{20}$) obtained using detrended fluctuation analysis with fourth order polynomial detrending. The vertical lines represent the standard deviation about the mean value of the local slopes for 25 realizations. The local slopes of the volatility series corresponding to Gaussian innovations (NORM) is shown (o) in each subplot as a reference. The horizontal solid lines in each subplot corresponds to scaling exponent of uncorrelated noise (α = 0.5).

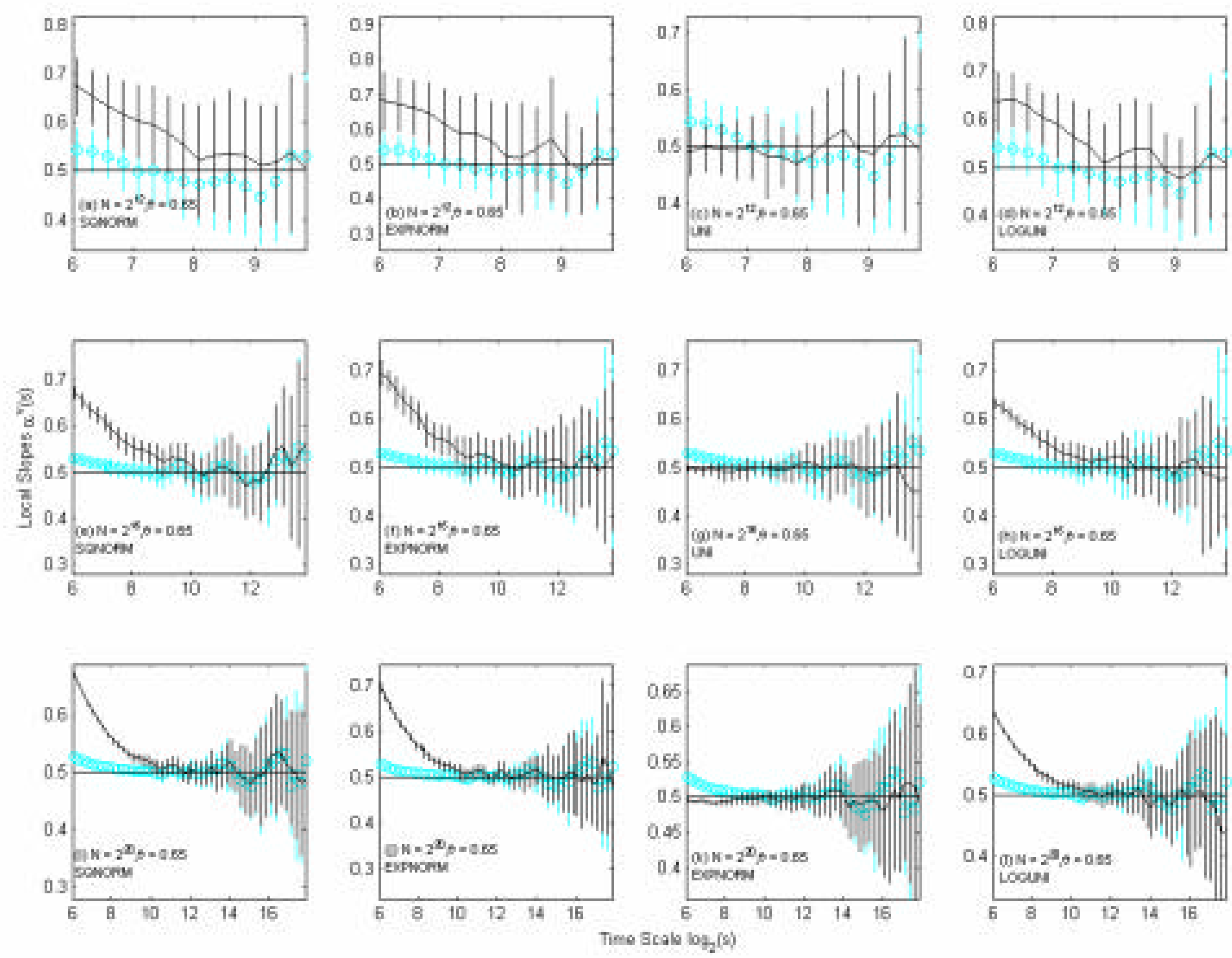

**Figure 6** Local slopes $\alpha(s)$ versus time scale $\log_2(s)$ of the volatility series of the stationary linear feedback process ($q$ = 0.65) with non-Gaussian innovations (SQNORM, EXPNORM, UNI, LOGUNI) and varying sample sizes (N = $2^{12}$, $2^{16}$ and $2^{20}$) obtained using detrended fluctuation analysis with fourth order polynomial detrending. The vertical lines represent the standard deviation about the mean value of the local slopes for 25 realizations. The local slopes of the volatility series corresponding to Gaussian innovations (NORM) is shown (o) in each subplot as a reference. The horizontal solid lines in each subplot corresponds to scaling exponent of uncorrelated noise ($\alpha$ = 0.5).